\documentclass[journal=jacsat,manuscript=letter]{achemso}

\usepackage[version=3]{mhchem} 
\usepackage{gensymb}
\usepackage{soul}
\usepackage{multirow}
\usepackage{hyperref}

\author{Sijia Ke}
\email{sijia_ke@berkeley.edu}
\affiliation[UCB MSE]
{Department of Materials Science and Engineering, University of California at Berkeley, California 94720, USA}
\alsoaffiliation[LBNL CSD]
{Chemical Sciences Division, Lawrence Berkeley National Laboratory, Berkeley, California 94720, USA}
\author{Yusuf Shaidu}
\affiliation[UCB PHYS]
{Department of Physics, University of California at Berkeley, California 94720, USA}
\author{Jeffrey B. Neaton}
\email{jbneaton@lbl.gov}
\affiliation[UCB PHYS]
{Department of Physics, University of California at Berkeley, California 94720, USA}
\alsoaffiliation[LBNL MSD]
{Materials Sciences Division, Lawrence Berkeley National Laboratory, Berkeley, California 94720, USA}

\title{Enhanced Interlayer Coupling and Excitons in Twin-Stacked Two-Dimensional Magnetic CrSBr Bilayers}

\begin{document}

\maketitle
\begin{abstract}
The degree of electronic coupling between individual layers in few-layer van der Waals heterostructures offers a route to engineer their magnetic, electronic, and optical functionalities. Using state-of-the-art first-principles calculations, we demonstrate that the electronic coupling between two monolayers of CrSBr — an anisotropic two-dimensional magnetic semiconductor — is highly nonlinear and nonmonotonic with respect to their relative twist angle, exhibiting a pronounced maximum at the twin-stacking configuration. The coupling strength scales with both the degree of overlap of Br orbitals adjacent to the van der Waals gap and the cosine of half of the interlayer spin angle. This enhanced interlayer electronic coupling gives rise to excitons delocalized across the two layers with a strong polarization dependence that reflects the details of the interlayer spin alignment. Our results reveal a sensitive interplay between twist angle, magnetism, and excitonic properties in twin-stacked CrSBr bilayers, and they establish twin stacking as an effective route to engineering interlayer coupling and optical response in anisotropic two-dimensional magnets with rectangular lattices.
\end{abstract}
Interlayer electronic coupling plays an essential role in layer-dependent electronic structure and optical properties in stacks of two-dimensional (2D) van der Waals (vdW) materials \cite{barre2024,tian2025}. In few-layer vdW heterostructures, orbital overlap across neighboring 2D monolayers can lead to electronic states delocalized across multiple layers, altering their energies and character in a way that is wavevector-dependent\cite{barre2024}. Consequently, 2D vdW materials often exhibit pronounced layer-dependent electronic structure and optical properties \cite{barre2024,qiu2017,naik2022}. However, the nature of this layer dependence varies widely among 2D vdW materials. For example, the optical gap of black phosphorous drops from 1.6 eV for the monolayer to 0.55 eV for four layers, reflecting strong interlayer electronic coupling \cite{qiu2017}. In contrast, CrSBr exhibits nearly thickness-independent optical gaps because its interlayer antiferromagnetic (AFM) configuration suppresses interlayer electronic coupling. However, it has been demonstrated that an external magnetic field can render CrSBr interlayer ferromagnetic (FM), thereby activating interlayer electronic coupling \cite{ziebel2024,cenker2022}. Such coupling delocalizes states across layers and modifies dielectric screening, with consequences for the electronic structure across the Brillouin zone. \par 
In addition to external fields, relative stacking and registry are also used to engineer interlayer electronic coupling. Twisting -- which alters local atomic stacking through a relative rotational angle -- leads to moiré potential landscapes that can drastically change both real and reciprocal space wavefunctions, inducing diverse emergent phenomena \cite{cao2016,cao2020,ciarrocchi2022,shaidu2025,naik2022,chen2024,zhang2025}. Extensive studies have explored twisted materials with hexagonal lattices \cite{liu2014,cao2016}, such as graphene and transition metal dichalcogenides, but twisting materials with rectangular lattices has remained largely underexplored. Recent work has demonstrated that rectangular lattices support a broad set of commensurate twist angles \cite{an2025,sun2025}. Among these angles, those close to the twin-stacking angle, where a diagonal of the rectangular lattice of the upper layer aligns with the opposite diagonal of the other layer, produce a 1D-like moiré pattern and host quasi-1D electronic behavior \cite{sun2025,an2025}. \par 
Very recently, experimental measurements on twisted CrSBr have suggested that strong in-plane magnetic anisotropy prevents any complex in-plane spin texture and conserves intralayer ferromagnetic order in twisted CrSBr, spanning from small ($\sim 3\degree$) to large twist angles \cite{mondal2025,sun2025,chen2024,li2025,zhang2025}. Within this regime, tunable exciton energies and linear dichroism with the twist angle have been experimentally observed \cite{li2025}; moreover, a recent first-principles study found an increase in interlayer distance in CrSBr bilayers upon twisting, which can decrease the interlayer magnetic interaction \cite{zhang2025}. \par 
In this letter, we use density functional theory (DFT) and the ab initio \textit{GW}-Bethe Salpeter equation (BSE) approach to investigate how large-angle twisting of CrSBr bilayers modulates the interlayer electronic coupling and affects the electronic and optical properties. We demonstrate a highly non-linear and nonmonotonic dependence of interlayer coupling on twist angle, governed by changes in both overlap between bromine \textit{p} orbitals on adjacent layers across the vdW gap and angles between spins in adjacent layers. In particular, the interlayer coupling is maximized at the twin-stacking angle, $2 \arctan(\frac{b}{a})$, where $a$ and $b$ are in-plane rectangular lattice parameters. We analyze the consequences of this enhanced interlayer coupling for the electronic structure of twisted CrSBr and perform GW-BSE calculations that reveal twist-enabled mixing of intralayer excitons, distinct from untwisted bilayers, and reflected in their polarization dependence. Our results elucidate how twin stacking couples twist angle, spin order, and excitons in rectangular-lattice two-dimensional magnets. \par
\begin{figure}
    \centering
    \includegraphics[width=1\linewidth]{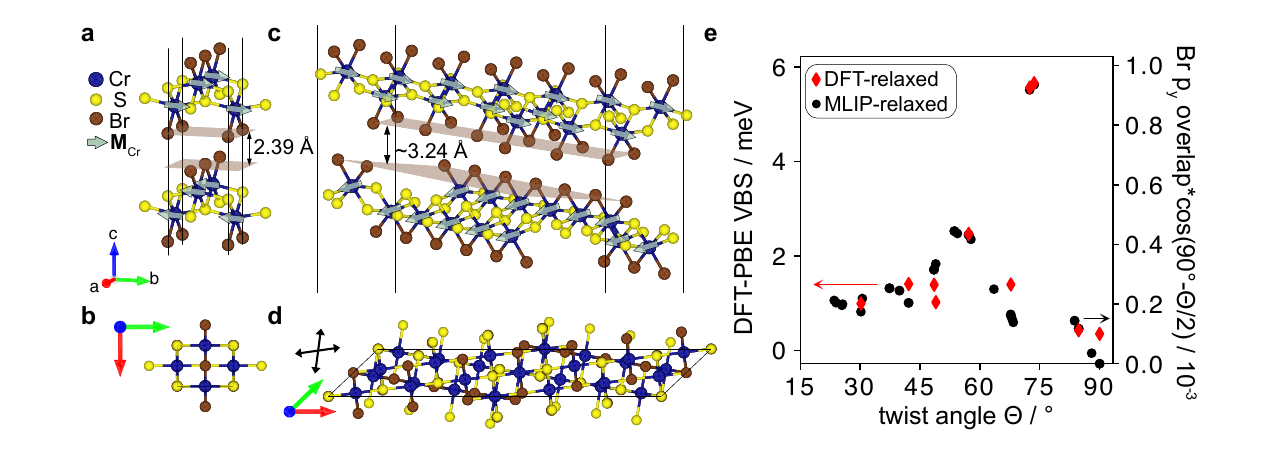}
    \caption{\textbf{a-d} The atomic structure of untwisted (\textbf{a, b}), near twin-stacked angle (\textbf{c, d}). The magnetic easy axis of each layer is shown as an black arrow for near twin-stacked angle in \textbf{d}. The local magnet moments are represented as green arrows on Cr. \textbf{e} DFT-PBE calculated valence band maximum splitting (VBS, red diamond), used as a measure of the interlayer electronic coupling, shown in the left axis, and the multiple of Br $p_{y}$-$p_{y}$ overlap (defined in SI) and cosine of half angle between spins (180$\degree - \Theta$) across MLIP-relaxed CrSBr bilayers (black dot), shown in the right axis, both as a function of twist angle. Interlayer electronic coupling is orbital forbidden in the 90$\degree$-twisted structure.}
    \label{fig:show}
\end{figure}

\begin{figure}
    \centering
    \includegraphics[width=1\linewidth]{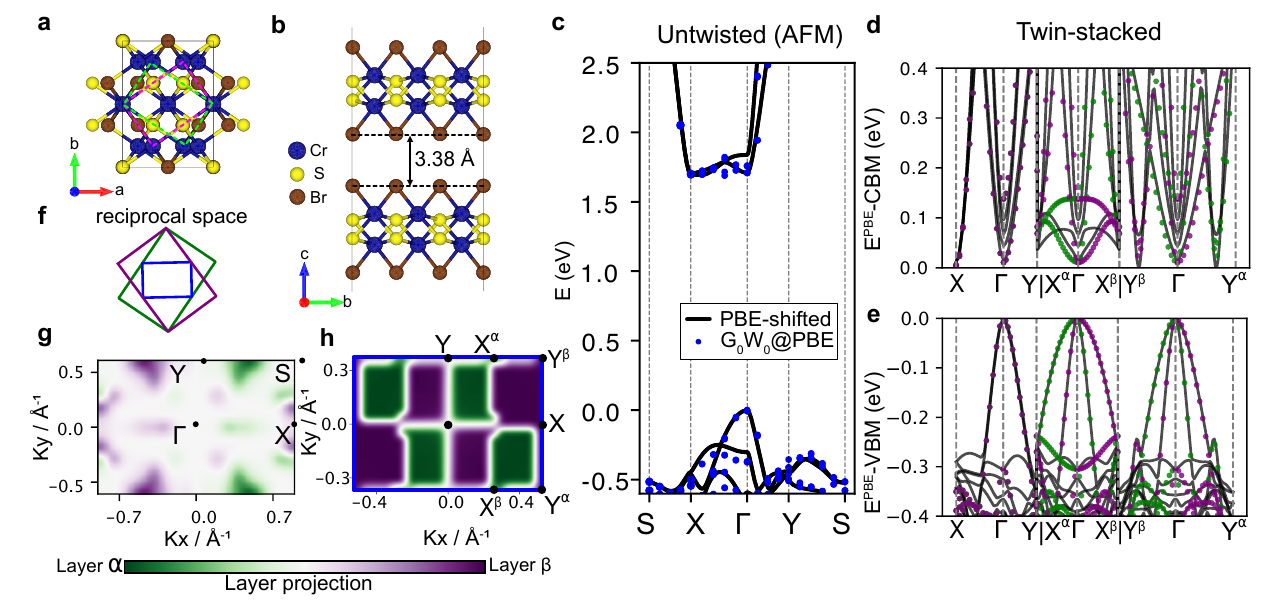}
    \caption{\textbf{a-b} Top (\textbf{a}) and side (\textbf{b}) view of a representative twin-stacked CrSBr bilayer. The green and purple dashed rectangles in (\textbf{a}) indicate the primitive unit cells of each monolayer before relaxation, and notably the diagonal of these two rectangles overlap. \textbf{c} $G_{0}W_{0}$@PBE and PBE calculated band structure of the untwisted CrSBr bilayer, with the reciprocal space points shown in \textbf{g}. Layer $\alpha$ is the lower layer and layer $\beta$ is the upper layer, shown for example in (\textbf{b}). \textbf{d, e} The conduction (\textbf{d}) and valence (\textbf{e}) band structure in the twin-stacked CrSBr supercell shown in black curves. The green and purple points represent states of the monolayer primitive cell folded into the Brillouin zones of twin-stacked CrSBr. The chosen reciprocal space paths and their ends are shown in \textbf{h}. \textbf{f} The purple and green solid rectangles represent the Brillouin zones of the monolayer primitive unit cell, and the blue solid rectangle represents the Brillouin zone of the twin-stacked CrSBr bilayer supercell.
    \textbf{g, h} Layer contributions to the lowest conduction band states for untwisted (\textbf{g}) and the twin-stacked (\textbf{h}) CrSBr.}
    \label{fig:electronic}
\end{figure}

\begin{figure}
    \centering
    \includegraphics[width=1\linewidth]{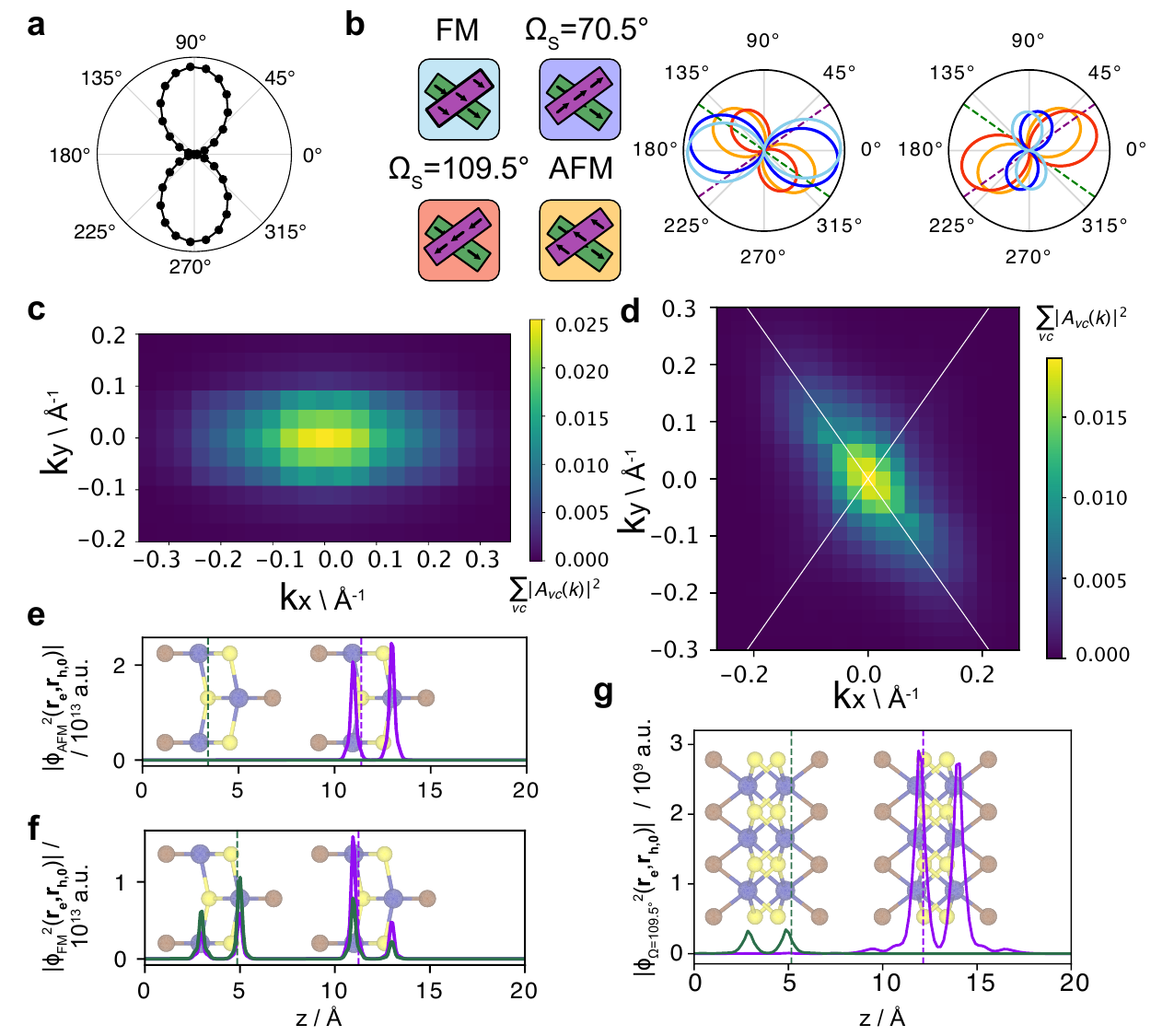}
    \caption{\textbf{a} Polarization dependence of the lowest-lying exciton in the untwisted CrSBr bilayers. \textbf{b} Polarization dependences of the two lowest-lying excitons in a 36-atom twisted CrSBr bilayer. The purple and green dashed lines on the polar plots indicate the magnetic easy axis of the two layers. The light blue curves are computed from the interlayer FM case, the darker blue curves are associated with the $\Omega_{S} =\Theta^{twin}$ case, the red curves are from our calculations with $\Omega_{S} = 180 \degree - \Theta^{twin}$, and the orange curves are from the interlayer AFM case. \textbf{c, d} The sum of exciton coefficients associated with transitions from occupied valence $v$ to unoccupied conduction $c$ states at wavevector $k$ ($\sum_{v,c}|A^{S}_{vck}|^{2}$) in the untwisted (\textbf{c}) and the twin-stacked (\textbf{d}) structure for the lowest exciton. \textbf{e-g} Magnitude of the first exciton wavefunction in solid curves for the untwisted bilayer with interlayer AFM order (\textbf{e}), interlayer FM order (\textbf{f}), and the twin-stacked bilayer with canted magnetic configuration $\Omega_{S} = 180 \degree - \Theta^{twin} = 109.5 \degree$ (\textbf{g}) along \textit{z}, normal to the bilayer with the hole position fixed in the upper and lower layer, respectively, shown in the dashed lines of color corresponding to the plotted wavefunction. (Green curves represent the lower layer and purple curves represent the upper layer.) A side view of the untwisted and twin-stacked CrSBr atomic structure is shown, aligned with \textit{z} coordinates for reference.}
    \label{fig:ex-all}
\end{figure}

 
The measured lattice parameters of the rectangular untwisted CrSBr bilayer (Fig. \ref{fig:show}a) \cite{beck1990} primarily used in this work are, a = 3.508 Å  and b = 4.763 Å, which differ by $\sim$ 25\% and give rise to strong in-plane anisotropy in spin orientation, electronic structure, and optical properties. CrSBr exhibits intralayer ferromagnetic (FM) order with a magnetic easy axis along \textit{b} and an interlayer antiferromagnetic (AFM) ground state (shown in Fig. \ref{fig:show}b). The band structure of CrSBr features significant anisotropy with weak dispersion along $\Gamma-\mathrm{X}$ and a large dispersion along $\Gamma-\mathrm{Y}$ for both the lowest conduction band and the highest valence band (Fig. \ref{fig:electronic}c). Consistent with this anisotropy, the first two bright excitons of CrSBr bilayer are computed to have a strong polarization dependence along the \textit{b}-axis (Fig. \ref{fig:ex-all}a). \par
To investigate the effect of twisting on interlayer coupling, we construct large CrSBr bilayer moiré supercells with twist angles, $\Theta$, ranging from 25$\degree$-90$\degree$ (see Supporting Information (SI) for details). We perform first-principles DFT calculations on several moiré supercells that contain fewer than 250 atoms corresponding to different $\Theta$. For those twist angles requiring larger unit cells, we use a fine-tuned machine-learned interatomic potential (MLIP) based on a state-of-the-art equivariant and message passing model, MACE \cite{batatia2025}, to relax the atomic structures. Our MACE foundation model \cite{batatia2025} is fine-tuned on relaxation trajectories of bilayer CrSBr supercells computed using vdW-corrected DFT-PBE methods (see SI for details). \par
While the interlayer electronic coupling of the untwisted CrSBr bilayer is suppressed by the ground-state interlayer antiparallel spin alignment, prior experiments for twisted CrSBr bilayers have suggested a noncollinear canted magnetic configuration while preserving intralayer FM, with the Cr spins aligned along the magnetic easy axis (\textit{b}-axis) in each layer \cite{li2025,mondal2025,sun2025}. As the bilayers are twisted relative to each other, the spins follow the easy axis associated with each individual layer; this leads to noncollinear magnetic configurations and nonzero interlayer electronic coupling in twisted CrSBr bilayers. The spin configurations in our calculations are relaxed from an initial input magnetic configuration, and spins are found to align with the easy axis of each layer such that the relative spin angle $\Omega_{S}$ between spins in each layer equals 180$\degree - \Theta$ (and is set to be larger than 90$\degree$, consistent with the angle between easy axes). Schematics of spin alignments are shown in Fig. \ref{fig:show}a,c.  \par
\begin{figure}
    \centering
    \includegraphics[width=0.5\linewidth]{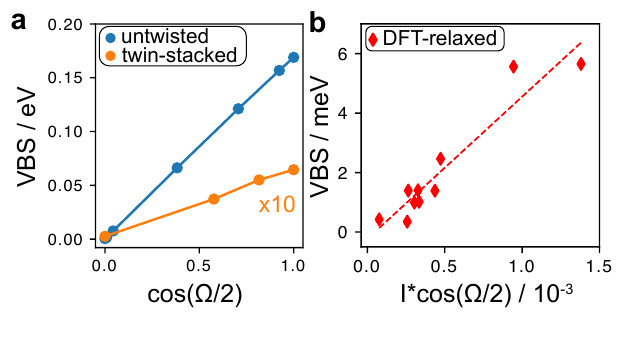}
    \caption{\textbf{a} Valence band maximum splitting (VBS) in untwisted (blue) and twin-stacked (orange) CrSBr bilayers as a function of the cosine of half the interlayer spin angle, where the twin-stacked VBS is increased by a factor of 10 for visibility. \textbf{b} VBS in several twisted CrSBr scales with the multiple of Br $p_{y}$-$p_{y}$ orbital overlap and cosine of half of the spin angle.}
    \label{fig:py}
\end{figure}

 
With noncollinear spin-alignment allowing interlayer electronic coupling, the valence band maximum (VBM) (or conduction band minimum, CBM) states at $\Gamma$ in the interlayer AFM case are split into two non-degenerate states in energy upon twisting. In this work, we use the valence band maxima splitting (VBS) as a measure of the magnitude of interlayer electronic coupling as a function of twist angle. (Using the conduction band splitting would lead to the same qualitative results.) In a tight-binding picture, the interlayer electronic hopping strength is proportional to the inner product between the spinor components of the monolayer orbitals, scaling as $\cos(\Omega_{S}/2)$, where $\Omega_{S}$ is the spin angle between layers, in agreement trend-wise with our calculated VBS in Fig. \ref{fig:py}a. \par
Fig. \ref{fig:show}e shows that the computed VBS of our large-angle twisted supercells changes nonlinearly and nonmonotonically with twist angle, exhibiting a pronounced peak near a specific angle, $\Theta^{twin} = 2 \arctan(\frac{b}{a})$, the twin-stacking angle. Since the VBM has predominantly Br $p_{y}$ orbital character (see SI Fig. S5) and the atoms adjacent to the vdW gap are Br, we expect the interlayer electronic coupling depends on the relative placement and orientation of the Br \textit{p} orbitals adjacent to the gap. \par
To capture this effect, we perform model calculations where we define the Br $p_{y}$-$p_{y}$ orbital overlap, $I$, as the sum of overlaps between two \textit{p}-type Gaussian functions located at Br adjacent to the vdW gap, explicitly accounting for the orbital orientations. Details of the overlap are provided in SI. Fig. \ref{fig:py}c shows the DFT-PBE calculated VBS scales with $I\cos(\Omega_{S}/2)$. Fig. \ref{fig:show}e shows the computed Br $p_{y}$-$p_{y}$ orbital overlap on our MLIP relaxed supercells as a function of twist angle, which likewise displays a strongly nonmonotonic behavior with a pronounced maximum near the twin-stacking angle. An \textit{s}-type Gaussian overlap is also tested, and as expected yields poor agreement with the calculated VBS (see SI Fig. S7). \par
Our relaxed moiré supercells exhibit atomic corrugations and an overall increase in the vdW interlayer distance by almost 1 Å (Fig. \ref{fig:electronic}b and Fig. \ref{fig:show}c) compared to the untwisted structure, reducing interlayer coupling. This increased interlayer distance has also been found in previous studies of twisted CrSBr bilayers \cite{zhang2025}.  The computed VBS in these supercells decreases by nearly two orders of magnitude compared with that in the untwisted case with interlayer FM.  \par
In Fig. S7 of the SI, we show Br $p_{y}$-$p_{y}$ orbital overlap evaluated without atomic corrugations and with a fixed interlayer distance of 3.3 Å. Remarkably, the pronounced peak at $\Theta^{twin}$ persists. This observation suggests that the enhanced interlayer coupling at $\Theta^{twin}$ arises primarily from in-plane registry of Br atoms, the Br $p_{y}$ VBM character, and the magnetic anisotropy, rather than structural corrugation effects alone. In the untwisted bilayer and the 90$\degree$ twisted limits, the interlayer electronic coupling vanishes, because the former exhibits interlayer AFM order whereas the latter features mutually perpendicular Br $p_{y}$ orbitals. \par

We find that the band structure of the twisted supercell largely resembles the superposition of the two pristine monolayer band structures folded into the reduced Brillouin zone of the twisted supercell. As a result, key features of monolayer CrSBr, such as the lowest transition at $\Gamma$, and the strongly anisotropic band curvatures, are preserved in the moiré supercells. We use a model 36-atom twisted CrSBr bilayer supercell to illustrate these additive features. The 36-atom supercell structure is constructed by stacking two supercell monolayers with equal strain and a relative rotation of $\Theta^{twin} = 70.53 \degree$, where lattice parameters of the monolayer primitive cell are changed to $a = $ 3.426 $\text{\AA}$ (2.3\% contraction) and $b = $ 4.846 $\text{\AA}$ (1.7\% expansion). Although this particular supercell construction involves a larger and uniform strain and the same strain sign in each layer compared to the approach described above, this supercell exhibits an orthorhombic lattice, and each layer contains only 18 atoms (Fig. \ref{fig:electronic} ab), which simplifies the band folding analysis and renders excited state calculations computationally feasible. \par
Fig. \ref{fig:electronic}a,f (green and purple curves) shows the atomic structure of the pristine monolayer unit cell highlighted and its corresponding first Brillouin zones. Fig. \ref{fig:electronic}d-e show that DFT-PBE calculated eigenvalues of monolayer CrSBr folded inside the Brillouin zone (scattered markers, with green and purple differentiating the two bilayers) largely overlap with DFT-PBE calculated conduction and valence band structure (black curves) of the twin-stacked structure. The DFT-PBE calculated VBS in twin-stacked structure is 4 meV with a relative spin angle of $180\degree - \Theta^{twin}$. Conduction states along the originally nearly flat $\Gamma$-$\text{X}$ line for the monolayer primitive cells are folded into $k_{y} = \pm  2 k_{x}$ for $|k_{x}| < 0.25$, $k_{y} = \pm(2k_{x}-0.5)$ for $k_{x} > 0.25$, and $k_{y} = \pm 2 (k_{x}+0.5)$ for $k_{x}<- 0.25$ with wavevectors expressed in crystal coordinates; see SI Fig. S8. Fig. \ref{fig:electronic}h shows the layer contribution of the lowest conduction states in reciprocal space, where most regions are dominated by one layer and limited regions are interlayer hybridized, for example states near $k_{y} = 0$ and  $k_{x} = 0$. This behavior contrasts with that of untwisted structure in the interlayer FM configuration (Fig. \ref{fig:electronic}g), where a much larger portion of reciprocal space exhibits interlayer hybridization. These differences have important implications for optical selection rules and excitonic properties, as discussed below. \par

We perform \textit{GW}-BSE calculations to study the excited state properties in untwisted and twin-stacked CrSBr. In an untwisted CrSBr bilayer with interlayer AFM, the lowest two excitons are bright for incident light polarized along the \textit{b} axis (Fig. \ref{fig:ex-all}a) and are degenerate in energy. These two excitons are each localized within a different layer and are 2D Wannier-Mott-like with computed exciton binding energies of 0.73 eV. In the twin-stacked CrSBr bilayer ($\Theta^{twin}$), the two lowest excitons are nearly degenerate in energy with a slightly reduced binding energy of 0.60 eV and a small residual splitting, and are bright with distinct polarization maxima aligned closely along the easy axis of each layer, respectively (Fig. \ref{fig:ex-all}b, red curves). This polarization behavior arises from the electron-hole interactions. In a single-particle picture that neglects electron-hole interactions, the two lowest transitions at $\Gamma$ feature states from both layers, leading to the polarization of these two direct transitions along either the bisectors of the two easy axes of both layers. \par
Fig. \ref{fig:ex-all}c and d shows the sum of the coefficients $\sum_{v,c}|A^{S}_{vc\textbf{k}}|^{2}$ of the lowest-lying $S=1$ exciton of the untwisted and twin-stacked bilayers resolved by wavevector $\textbf{k}$. (Each coefficient $A^{S}_{vc\textbf{k}}$ describes the contributions of transitions from occupied valence states to unoccupied conduction states for a give $\textbf{k}$ for excited state $S$.) In the untwisted CrSBr bilayer, the first exciton is centered on the $\Gamma-\text{X}$ path where the band transition energies are smaller (Fig. \ref{fig:ex-all}c). By contrast, in the twin-stacked CrSBr bilayer, the exciton is centered near $k_{y} = -2 k_{x}$ ($|k^{AA}_{x}| < 0.25$), where the highest occupied and lowest unoccupied states are dominated by the upper layer (Fig. \ref{fig:ex-all}d), reflecting the alignment of this path with the $\Gamma-\text{X}$ direction of the upper monolayer. \par

Although the first two bright excitons in twin-stacked CrSBr bilayers retain intralayer character, they also exhibit finite mixing with the opposite layer, a consequence of the twisting-enabled interlayer coupling. Fig. \ref{fig:ex-all}g shows the first exciton wavefunction when the hole position is fixed in the lower (green curve) and upper (purple curve) layers. Notably, while the exciton wavefunction of the twisted bilayer remains largely distributed within the same layer as the fixed hole position, it now exhibits a sizable amplitude in the opposite layer, in contrast to the case of the untwisted CrSBr bilayer (green and purple curves comparison in Fig. \ref{fig:ex-all}e). \par
We further study the effect of interlayer electronic coupling by altering the relative alignment between the spins in the two layers. In untwisted CrSBr bilayer with interlayer ferromagnetic order, the strong interlayer electronic coupling mixes the two lowest excitons, breaking their energy degeneracy and rendering the interlayer hybrid exciton (Fig. \ref{fig:ex-all}f) with the lowest exciton binding energy of 0.71 eV. Our computed exciton energies in this configuration are in qualitative agreement with prior calculations and experiments \cite{wilson2021}. \par
In the twin-stacked CrSBr bilayer, the alignment of the transition dipole polarization of the two lowest excitons depends on the relative alignment of the spins in the two layers. Fig. \ref{fig:ex-all}b shows the polarization dependence of the transition dipoles of the lowest two excitons for three representative spin alignments: interlayer FM, noncollinear ($\Omega_{S} \approx \Theta^{twin}$ and $\Omega_{S} \approx 180\degree -\Theta^{twin}$), and interlayer AFM. In the interlayer AFM case (Fig. \ref{fig:ex-all}b orange), the transition dipoles of the excitons align with the easy axis of each layer, reflecting suppressed interlayer electronic coupling and predominantly intralayer excitons. In other magnetic configurations, interlayer electronic coupling mixes the two excitons, and the stronger the coupling, the larger the deviation of their transition dipoles from the respective easy axes. \par 
We note that the effect of magnetic configuration-induced interlayer electronic coupling on the excitons of twin-stacked CrSBr differs qualitatively from the untwisted bilayer in the interlayer FM order because of the distinct momentum-space character of the interlayer electronic coupling in the two structures. The strong interlayer electronic coupling in untwisted interlayer FM CrSBr bilayers is associated with nearly complete hybridization of single-particle states across the Brillouin zone (Fig. \ref{fig:electronic}g); in contrast, the weaker interlayer electronic coupling in the twin-stacked structure arises from partial hybridization in limited regions of reciprocal space, together with increased interlayer spacing (Fig. \ref{fig:electronic}h). In both structures, the interdependence of interlayer magnetism and excitons is a signature of strong magnon-exciton coupling, which is consistent with prior experimental measurements \cite{sun2024,sun2025}. \par

In addition, the changes in transition dipole dependence with relative interlayer spin alignment suggest a potential optical probe that uses photoluminescence polarization to identify interlayer magnetic configurations in ultrathin few-layer CrSBr, systems for which standard magnetometry techniques tend to fail \cite{burch2018}. We expect similar polarization dependence in other moiré configurations, with the polarization of the two excitons aligning closely with the easy axis of each layer, respectively. Deviation from each easy axis depends on the strength of interlayer electronic coupling-induced mixing between the two intralayer excitons. Although the two excitons are nearly degenerate in energy in our calculations, differences in substrate dielectric environment or strain could increase their splitting, beyond what is calculated here, facilitating experimental resolution of the two excitons in polarization-resolved measurements, as has been observed in prior work \cite{sun2025}. \par

In this work, we demonstrate how twisting introduces changes in structures and interlayer magnetic configuration, affecting interlayer electronic coupling, electronic structure, spin ordering, and excitonic properties in untwisted and twisted magnetic CrSBr bilayers. We find that the interlayer coupling reaches a pronounced maximum near a specific twist angle, the twin-stacking angle. We attribute the nonmonotonic and nonlinear trends in coupling with twist angles to the overlap between Br $p_{y}$ orbitals adjacent to the vdW gap multiplied by the cosine of half the interlayer spin angles, a relationship supported by our detailed first-principles calculations. Although twisting increases the average interlayer spacing, it simultaneously enables greater orbital overlap and a noncollinear interlayer magnetic configuration and finite interlayer electronic coupling. Moreover, twisting reshapes electronic structure such that only a fraction of states in reciprocal space become interlayer hybridized. Our GW-BSE calculations for the twin-stacking structure show that the two lowest bright excitons exhibit polarizations closely aligned with the easy axis of each layer, with deviations governed by the strength of interlayer electronic coupling. Our results suggest a new optical route to identify interlayer magnetic configurations in ultrathin CrSBr and related anisotropic two-dimensional magnets. More broadly, our work highlights how stacking geometry and magnetism can jointly engineer interlayer coupling and excitons in twisted vdW magnets at the intersection of twistronics and spintronics. \par 
\section{Methods}
We use Quantum ESPRESSO \cite{giannozzi2009a,giannozzi2017a} and optimized norm-conserving pseudopotentials (ONCVPSP v0.4) from PseudoDojo \cite{hamann2013,vansetten2018a}. We use density functional theory (DFT) within the generalized gradient approximation of Perdew, Burke, and Ernzerhof (PBE) \cite{perdew1996b} and we account for dispersion interactions using Grimme's DFT-D3 van der Waals corrections with Becke and Johnson damping \cite{grimme2010a} for all calculations. We use a plane wave kinetic energy cutoff of 94\,Ry for all calculations. Total energies are converged to within 2$\times 10^{-7}$ eV/atom, and all Hellmann-Feynman forces are converged to below 3$\times 10^{-2}$ eV/$\text{\AA}$. We fix the in-plane lattice parameters to match the experimental bulk values, a = 3.508\,$\text{\AA}$, b = 4.763\,$\text{\AA}$, and the out-of-plane lattice parameter of our supercell is set to 30\,$\text{\AA}$ for the untwisted CrSBr bilayer, which is further used to construct twisted CrSBr bilayers. Structural relaxations are performed using scalar spin calculations, while electronic structure calculations employ noncollinear magnetism with spin-orbit coupling included.\par
We use the twist\_layers code \cite{shaidu2025} to construct twisted CrSBr bilayers at different twist angles. The details of twist\_layers code are discussed in SI. We perform structural relaxation for twisted CrSBr bilayers at twist angles of 30.2$\degree$, 42.2$\degree$, 48.5$\degree$, 57.2$\degree$, 67.8$\degree$, 72.5$\degree$, 73.6$\degree$, 84$\degree$, 90$\degree$. The relaxation energies and forces from these twisted CrSBr bilayers, together with slid untwisted CrSBr bilayers (see details in SI), are used to fine-tune an equivariant machine learning interatomic potential foundation model based on the MACE architecture (MACE-mp0) \cite{batatia2025}. The fine-tuned MLIP is further used to relax other twisted atomic structures containing a larger number of atoms. \par 
We use BerkeleyGW \cite{deslippe2012} to perform one-shot $G_{0}W_{0}$-BSE calculations using a PBE starting point with noncollinear configurations. We compute the inverse dielectric matrix with the frequency dependence modeled using the Godby-Needs generalized plasmon pole model \cite{deslippe2013,godby1989,oschlies1995}. We use a slab Coulomb truncation to reduce interactions between periodic images along the \textit{z} direction \cite{ismail-beigi2006}. Nonuniform neck subsampling is used to construct the dielectric matrix \cite{dajornada2017a}. We include the static remainder \cite{deslippe2013} when computing self energies. We note that time-reversal symmetry, which is assumed in standard Alder-Wiser formalism for the polarizability \cite{adler1962,wiser1963} implemented in BerkeleyGW, is broken for noncollinear magnetic systems. Accordingly, we use the full k-point grid without symmetry reduction in computing the polarizability in our \textit{GW}-BSE calculations, and demonstrate that the standard Alder-Wiser formalism introduces only marginal differences for CrSBr bilayers.  \par 
For untwisted CrSBr bilayers, the inverse dielectric matrix $\varepsilon^{-1}_{\textbf{GG'}}(\textbf{q},w)$ is sampled on a full uniform $\Gamma$-centered $10\times8\times1$ k-mesh, and constructed using 1440 bands and a 20 Ry cutoff, for which convergence to 0.15 eV is reached. The BSE matrix elements are computed on a uniform $\Gamma$-centered $10\times8\times1$ coarse k-grid with 10 valence bands and 4 conduction bands, and then interpolated onto a rectangular patch of $50\times30\times1$ centered around $\Gamma$ point \cite{alvertis2023} with 6 valence bands and 4 conduction bands. 
For the 36-atom twisted CrSBr structure, the inverse dielectric matrix $\varepsilon^{-1}_{\textbf{GG'}}(\textbf{q},w)$ is sampled on a full uniform $\Gamma$-centered $8\times6\times1$ k-mesh, and constructed using 1500 bands and a 20 Ry cutoff, converging energies to 0.05 eV. The BSE matrix elements are computed on a uniform $\Gamma$-centered $8\times6\times1$ coarse k-grid with 14 valence bands and 4 conduction bands, then interpolated onto a rectangular patch of $40\times30\times1$ centered around $\Gamma$ point \cite{alvertis2023} with 14 valence bands and 4 conduction bands. We apply a scissor shift to correct quasiparticle transition energies, where valence band eigenvalues are reduced by -0.940 eV and conduction band eigenvalues are increased by 0.220 eV based on our $\textit{G}_{0}\textit{W}_{0}$@PBE calculations. This approach preserves smooth anisotropic band dispersions near $\Gamma$ that are essential for accurately describing excitons. (The rationale for performing scissor shifts is in SI.) We use the momentum operator to construct all dipole transition matrix elements.

\section{Acknowledgments}
The authors thank Yue Sun, Stephen Gant, and Mit Naik for helpful discussions. This work is supported by the Theory of Materials FWP supported by the U.S. Department of Energy, Office of Science, Office of Basic Energy Sciences, under Award No. DE-SC0020129. Computational resources provided by the National Energy Research Scientific Computing Center (NERSC), supported by the Office of Science of the Department of Energy operated under Contract No. DE-AC02-05CH11231 using NERSC Award No. ERCAP0033609.

\bibliography{crsbr-update.bib}

\end{document}